\documentclass[notitlepage,superscriptaddress,showpacs,nobalancelastpage,twocolumn,aps,prl]{revtex4-2}
\usepackage[english]{babel}
\RequirePackage[T1]{fontenc}
\RequirePackage{times} % favourite for note.

\usepackage{siunitx} % typesets numbers with units very nicely
\usepackage[italicdiff]{physics}
\usepackage{amsfonts}
\usepackage{lipsum}
\usepackage{multirow}
\usepackage{subfigure}
\usepackage{amsmath}
\usepackage{amssymb}
\usepackage{graphicx, epstopdf}
\usepackage{wasysym}
\usepackage{xspace}
\usepackage{array}
\usepackage[hidelinks]{hyperref}
\usepackage[dvipsnames]{xcolor}
\usepackage{my_symbols}
\usepackage{CJK}
\usepackage{bm}
\usepackage{verbatim}
\usepackage{tikz}
\usepackage{comment}
\usepackage{marginnote}
\usepackage[textwidth=1.5cm]{todonotes}
\usepackage[normalem]{ulem}
\usepackage{soul}
\usepackage{booktabs}
\usepackage[commandnameprefix=ifneeded]{changes}

\sethlcolor{green}

\newcounter{mycomment}

\excludecomment{comment}  %do not show comments

\begin{document}

\begin{CJK*}{UTF8}{gbsn} % Use default fonts from CJK (see below)
\title{Interpreting S-Parameter Spectra in Coupled Resonant Systems: The Role of Probing Configurations}
\author{Jiongjie Wang}
\affiliation{Department of Physics and State Key Laboratory of Surface Physics, Fudan University, Shanghai 200433, China}
\author{Jiang Xiao (萧江)}
\email{xiaojiang@fudan.edu.cn}
%\email[Corresponding author:~]{xiaojiang@fudan.edu.cn}
\affiliation{Department of Physics and State Key Laboratory of Surface Physics, Fudan University, Shanghai 200433, China}
\affiliation{Institute for Nanoelectronics Devices and Quantum Computing, Fudan University, Shanghai 200433, China}
\affiliation{Shanghai Research Center for Quantum Sciences, Shanghai 201315, China}
\affiliation{Shanghai Branch, Hefei National Laboratory, Shanghai 201315, China}
%\affiliation{Zhangjiang Fudan International Innovation Center, Fudan University, Shanghai 201210, China}

\begin{abstract}

The S-parameter $S_{21}$ is widely used to characterize the resonant properties of various systems. However, we demonstrates that the reliability of $S_{21}$ as a true indicator of system resonances depends heavily on the specific probing setup employed. While point-probe or weak probes preserve the integrity of the $S_{21}$ spectrum and accurately reflect system resonances, multi-point and strong probes can introduce significant discrepancies. These discrepancies can manifest as misleading features such as repulsive level anti-crossing or attractive level crossing, even in systems that are fundamentally uncoupled. This highlights the critical importance of selecting appropriate probing techniques to ensure a precise evaluation of system resonances. 

\end{abstract}

\maketitle
\end{CJK*}

%\section{Introduction}

\emph{Introduction -} 
The transmission coefficient $S_{21}$ is pivotal in analyzing resonances across diverse scientific and engineering systems. By precisely measuring the amplitude and phase of the transmitted signal, researchers can extract critical insights into the fundamental properties such as resonance frequencies, quality factors, and coupling efficiencies. This coefficient, typically determined through vector network analysis, greatly aids in understanding the complex interactions within systems and devices.

One primary method to investigate light-matter interactions involves placing materials within a cavity to assess how their interaction affects light's transmission and reflection probabilities. This approach has led to the development of specialized fields such as cavity quantum electrodynamics \cite{walther_cavity_2006,xu_quantum_2023}, cavity optomechanics \cite{fan_sharp_2002,maleki_tunable_2004,aspelmeyer_cavity_2014}, and cavity spintronics \cite{lachance-quirion_hybrid_2019, zare_rameshti_cavity_2022}. 
In these hybrid systems, $S_{21}$ can reveal a range of phenomena, particularly in cavity-magnetic systems \cite{auld_coupling_1963}, where the interaction between photons and magnons has prompted discussions on phenomena like repulsive level anti-crossing for coherent coupling \cite{zhang_strongly_2014,wang_bistability_2018}, the level attraction for dissipative coupling \cite{yang_anti-_2017,wang_dissipative_2020,yang_control_2019,li_coherent_2022}, exceptional point \cite{zhang_observation_2017,lai_observation_2019,zhang_exceptional_2024}, the electromagnetic induced transparency \cite{boller_observation_1991,artoni_electromagnetic_2016,qian_non-hermitian_2023}, Fano resonance \cite{miroshnichenko_nonlinear_2005,miroshnichenko_fano_2010} and bound state in continuum \cite{han_bound_2023} \etc.

However, analyzing $S_{21}$ in hybrid cavity-matter systems becomes challenging, especially when the probing channel also mediates the coupling. A notable example is seen in cavity-magnet systems, where internal modes interact among themselves and with the probing light  \cite{yao_microscopic_2019}, involving multiple modes at various locations. This complicates the separation of the probing channel from the system, raising concerns about the reliability of $S_{21}$ spectra as a true indicator of system properties.

The reliable intepretation of the reflection and transmission spectra of light in hybrid systems relies on a comprehensive theoretical framework. The input-output theory \cite{collett_squeezing_1984, garrison_quantum_2008, walls_quantum_2008} has traditionally been employed to derive analytical expressions for the transmission spectra. The input-output theory is very successful, but it becomes increasingly more complicated when multiple modes are involved. In this context, an equivalent but much simpler loop-theory developed by Yuan \etal \cite{yuan_loop_2020} allows for the straightforward formulation of the $S_{21}$ expression by analyzing a few diagrams, thereby greatly simplifying the input-output formalism. 
%Consequently, we have chosen to utilize loop-theory for the computation of transmission spectra in the current analysis presented in this Letter.

In this Letter, we delve into the reliability of the measured $S_{21}$ spectrum as an indicator of system resonances. Our investigation, including both anyalytical results from the loop theory (or input-output theory) and the simulation results from COMSOL Multiphysics, reveals that while there is typically a correlation between the $S_{21}$ spectrum and the resonant behaviors of the system, discrepancies arise when multiple internal modes of the system interact concurrently with the probing channel. In such scenarios, the transmission spectra may not accurately represent the true resonance characteristics of the system. This finding underscores the complexity of interpreting $S_{21}$ measurements in systems with multiple interacting modes.

%In this Letter, we aim to explore the extent to which the $S_{21}$ spectrum accurately reflects the resonances of the system. Our findings suggest that the correlation between the $S_{21}$ spectrum and the system resonances generally holds, however it may break down when the multiple internal modes of the system are coupled with the probing channel simultaneously. Under these conditions, the features in the transmission spectra do not align with the resonance characteristics of the system with or without the probing channel. 
%Thus, it becomes evident that an open channel cannot effectively serve dual roles in both probing and coupling, highlighting a fundamental limitation in using $S_{21}$ spectra for detailed system analysis under such complex conditions.

%\section{Model}

\begin{figure}[t]
  \includegraphics[width=0.8\columnwidth]{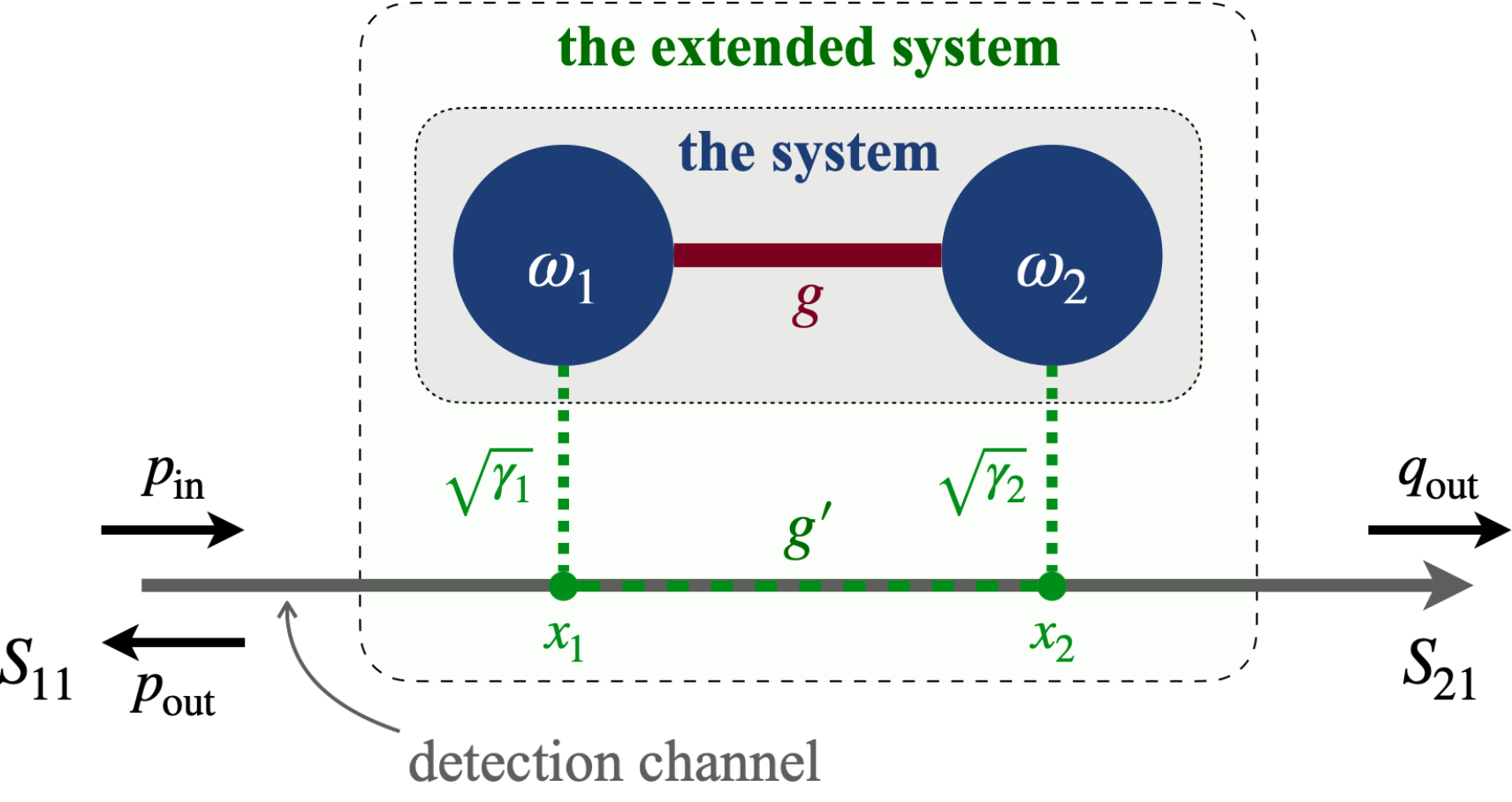}
  \caption{The setup for probing the resonances of a system with two coupled modes using the S-parameters. The two modes are in contact with the probing channel with strength $\sqrt{\gamma_i}$. The two modes have two distinct coupling routes: the direct coupling $g$ and the indirect coupling $g'\propto \sqrt{\gamma_1\gamma_2^*}$ via the probing channel.}
  \label{fig:model}
\end{figure}

\emph{Model -} To illustrate the correspondence between the resonances and the S-parameters, we use a simplified model as shown in \Figure{fig:model} consisting of two coupled modes connected to the probing channel, through which the reflection $S_{11}$ and transmission $S_{21}$ are measured. The overall Hamiltonian can be decomposed into the Hamiltonian for the system $\hH_0$, the open probing channel $\hH_c$, and the interaction between the system and the channel: $\hH = \hH_0 + \hH_c + \hH_I$ with 
%\wjj{In order to make the Hamiltonian here more closely resemble loop theory, a coupling marked with $\sqrt{\gamma^*}$ is adopted, rather than $\sqrt{\gamma_i}$. The subsequent derivation is based on the assumption that we take \(\sqrt{\gamma_i^*}\) here, so no modifications are necessary.}
\begin{subequations}
\label{eqn:H}
\begin{align}
%  \hH &= \hH_0 + \hH_c + \hH_I \qwith \\
  \hH_0 &= \hbar\omega_1 \ha_1^\dagger\ha_1 
  + \hbar\omega_2 \ha_1^\dagger\ha_2 
  + \hbar g \qty(\ha_1^\dagger\ha_2 + \ha_1\ha_2^\dagger), \\
  \hH_c &= \int \dd{\omega} \hbar\omega \qty[\hc_+^\dagger(\omega)\hc_+(\omega) + \hc_-^\dagger(\omega)\hc_-(\omega)], \\
  \hH_I &= \sum_{i=1}^2 \int \dd{\omega} \sqrt{\gamma_i^*}\qty[\ha_i^\dagger \hc_+(\omega)e^{i\phi_i} + \ha_i^\dagger \hc_-(\omega)e^{-i\phi_i}] + \mbox{h.c},
\end{align}
\end{subequations}
Here, $\omega_{1,2}$ are the natural (complex) frequencies of the individual modes whose imaginary parts represent the intrinsic dissipation of the mode, $g$ is the direct coupling between the two modes, and $\sqrt{\gamma_i}$ is the strength of the coupling with the probing channel. 
What's important here is distinguishing the coupling to the left- ($\hc_-$) and right-going ($\hc_+$) wave in the probing channel. Because the two modes may be spatially separated, thus the two modes couple to the probing wave with different phase $\phi_i$.
%Another feature that is mostly ignored in previous studies is the extra coupling $g'$ between the two modes via the probing channel, in addition to the direct coupling $g$. Furthermore, the coupling via the open channel can also rely on the spatial separation between the two oscillators, \ie the distance $l = \abs{x_2-x_1}$. This extra ingredient brings complications in infer the system resonances from the S-parameter measurement. The contact with open channel also brings extra dissipation to the modes, which shall be proportional to $\sqrt{\gamma_i}$. 

The resonant frequencies of the system corresponds to the real part of the eigenvalues of the following matrix
\begin{equation}
  \label{eqn:M0}
  W
%  = \mqty(\omega_1 & g_{12} \\ g_{21} & \omega_2)
  = \mqty(\omega_1 & g_\ssf{eff} \\ g_\ssf{eff} & \omega_2),
\end{equation}
where $g_\ssf{eff}$ is the effective coupling between the two modes. Depending on the concrete realization of coupling $g$ and the probing line, $g_\ssf{eff}$ can take any complex value. 
The real $g_\ssf{eff}$, corresponding to a coherent coupling, gives rise to the standard anti-crossing or level repulsive spectrum. For a purely imaginary $g_\ssf{eff}$, $W$ is a non-Hermitian matrix, and the spectrum exhibits a level attractive behavior. 
%The non-Hermitian type behavior can be found when the two modes are engineered to be a gain-loss system.
In the absence of the probing line, $g_\ssf{eff} = g$. 
%The value of $g$ actually gives the difference between the two eigenvalues at zero detuning ($\omega_1 = \omega_2$): $g = (\omega_+ - \omega_-)/2$. 

%This difference represented by $g$ is usually appears as the gap in the resonant spectrum.  When $g$ is real, the coupling between the two modes are coherent and the spectrum shall exhibit anti-crossing, or level repulsive behavior, \ie $\omega_\pm$ differ in their real parts. On the other hand, if $g$ is purely imaginary, the coupling becomes dissipative-like, and the spectrum exhibits a level attractive behavior, \ie $\omega_\pm$ have the same real part but differ in their imaginary parts. For a complex value of $g$, the spectrum has a mixed repulsive-attractive behavior.

%\section{Transmission and Absorption Spectrum}

\emph{Transmission -} By applying the loop theory proposed by Yuan \etal \cite{yuan_loop_2020}, an equivalent but much simpler theory to the input-output theory, we find the transmission coefficient for the two-mode model in \Figure{fig:model} as (see Appdenix)
\begin{equation}
  \label{eqn:S21}
  S_{21} = 1 + i\frac{|\gamma_1|\Delta_1 + \abs{\gamma_2}\Delta_2 
  - i\qty(\kappa_{12} g'_{21} - \kappa_{21}{g'}_{21}^*)\Delta_1\Delta_2}{1 - \kappa_{12}\kappa_{21}\Delta_1\Delta_2} 
\end{equation}
% \begin{subequations}
%   \label{eqn:S21}
%   \begin{alignat}{2}
%   S_{21} &= 1 + i\frac{|\gamma_1|\Delta_1 + \abs{\gamma_2}\Delta_2 
%   - i\qty(\kappa_{12} g'_{21} - \kappa_{21}{g'}_{21}^*)\Delta_1\Delta_2}{1 - \kappa_{12}\kappa_{21}\Delta_1\Delta_2} \\
%   S_{11} &= i\frac{|\gamma_1|\Delta_1e^{-iql} + \abs{\gamma_2}\Delta_2e^{iql}}{1 - \kappa_{12}\kappa_{21}\Delta_1\Delta_2} \nn
%   &\quad\quad\quad + \frac{\qty(\kappa_{12} g'_{21}e^{-iql} - \kappa_{21}{g'}_{21}^*e^{iql})\Delta_1\Delta_2}{1 - \kappa_{12}\kappa_{21}\Delta_1\Delta_2},
%   \end{alignat}
% \end{subequations}
where $1/\Delta_i = \omega - \omega_i + i|\gamma_i|$ is the detuning and $\kappa_{ij} = g_{ij} + g'_{ij}$ is the overall coupling with $g'_{ij} = -i \sqrt{\gamma_i^*\gamma_j}e^{iql}$ the extra coupling via the probing channel. Without loss of generality, we may assume $g'_{12} = g'_{21} = g'$ takes a common complex value. Here $ql = \phi_2 - \phi_1$ because the two modes are coupled with the same propagating wave in the channel with phase delay of $ql$. 
The minimum or the zeros of the transmission $S_{21}$ is often used to infer the resonant frequency of the system. The frequencies for these zeros are given by the eigenvalues of the $W$ matrix but with $g_\ssf{eff} = g_S$ and
% \begin{equation}
%   \label{eqn:MS}
%   M_S = \mqty(\omega_1 & g_S \\ g_S & \omega_2),
% %  (\omega_S - \omega_1)(\omega_S - \omega_2) = g_S^2,
% \end{equation}
% with the effective coupling 
\begin{equation}
  \label{eqn:gS}
    g_S^2 = g_{12}g_{21} + g_{21}(g'_{12} + {g'}_{21}^*)
    = g^2 + g(g' + g'^*).
\end{equation}
Similar to \Eq{eqn:M0}, $g_S$ represents the gap in the $S_{21}$ spectrum.
We see that the spectrum from $S_{21}$ matches with the spectrum of the system only when $\Re{g'} = 0$, \ie the probing channel does not route any extra coherent coupling. Such correspondence fails as long as $\Re{g'} \neq 0$, \ie when both modes couple to the probing channel ($\gamma_{1,2} \neq 0$) and the spatial separation between contacting points $l = \abs{x_2 - x_1} \neq n\pi$. 

%In addition to the probing role of the probing channel, the probing channel may play an active role in inducing extra coupling between the internal modes of the system. 

\emph{Absorption -} The absorption spectrum provides more accurate information about the resonances of the system. The absorption spectrum can also be obtained from S-parameters: $A(\omega) = 1- (\abs{S_{11}}^2 + \abs{S_{21}}^2)$. $A(\omega)$ maximizes at the system resonance frequencies, which are given by the eigenvalues of $W$ with $g_\ssf{eff} = g_A$ and
% \begin{equation}
%   \label{eqn:MA}
%   M_A = \mqty(\omega_1 & g_A \\ g_A & \omega_2),
% %  (\omega_A - \omega_1)(\omega_A - \omega_2) = g_A^2
% \end{equation}
% where the effective coupling
\begin{equation}
  \label{eqn:gA}
  g_A^2 = \kappa_{12} \kappa_{21} 
  = (g_{12} + g'_{12})(g_{21} + g'_{21})
  = (g+g')^2.
\end{equation}
The effective coupling $g_A$ simply include both the direct coupling $g$ and the indirect coupling $g'$ via the probing channel. Similarly, $g_A$ represents the gap in the absorption spectrum. The absorption spectrum matches with the spectrum of the isolated system when $g' =0$, \ie the probing channel does not provide any additional coupling. 
%This can be realized by making the probing channel a point contact with $\sqrt{\gamma_2} = 0$ for example, or the probing is very weak such that the indirect coupling is negligible: $\abs{g'} \ll \abs{g}$.

%Therefore, $g, g_S, g_A$ represent the gap of the absorption spectrum for the isolated system, the transmission spectrum, and the absorption spectrum of the extended system. 

% From \Eqs{eqn:gS}{eqn:gA}, we see that, $g_S$ and $g_A$, representing the gap of the transmission spectrum $S_{21}$ and the absorption spectrum, are both related to but different from the system gap $g$, in particular when the coupling via the probing channel cannot be neglected. Therefore, to infer the system spectrum from the transmission or absorption needs to a careful analysis.

% \begin{center}
% \begin{table}[t]
% \begin{tabular}{c|c|c} \hline
%   & $S_{21}$ Spectrum & Absorption Spectrum \\ \hline \hline
%   Point contact &  system & system \\ \hline
%   Multi weak contacts & system & system \\ \hline
%   Multi strong contacts & $\times$ & extended system \\ \hline
% \end{tabular}
% \label{tab:spectrum}
% \caption{The applicability of the transmission and absorption spectra in different detection setups.}
% \end{table}
% \end{center}

Both $S_{21}$ and the absorption spectrum are capable of providing genuine information about the system when the extra coupling channeled by the probing channel $g'$ vanishes or is negligible. This can be achieved by either i) making the probe to be point-contact, \ie only one $\gamma_i$ is non-zero so that $g' \propto \sqrt{\gamma_1^*\gamma_2} = 0$ or ii) the probe is multi-contact with $\gamma_{1,2} \neq 0$ but both probes are weak so that the indirect coupling is negligible: $\abs{g'} = \abs{\sqrt{\gamma_1^*\gamma_2}} \ll \abs{g}$. However, if the probe is multi-contact and strong such that $\abs{g'} \not \ll \abs{g}$, the transmission spectrum and the absorption spectrum do not match, and the $S_{21}$ spectrum could potentially mislead in understanding the (extended) system for such scenarios.

%\section{Micro-ring system simulation}

\begin{figure}[b]
  \includegraphics[width=\columnwidth]{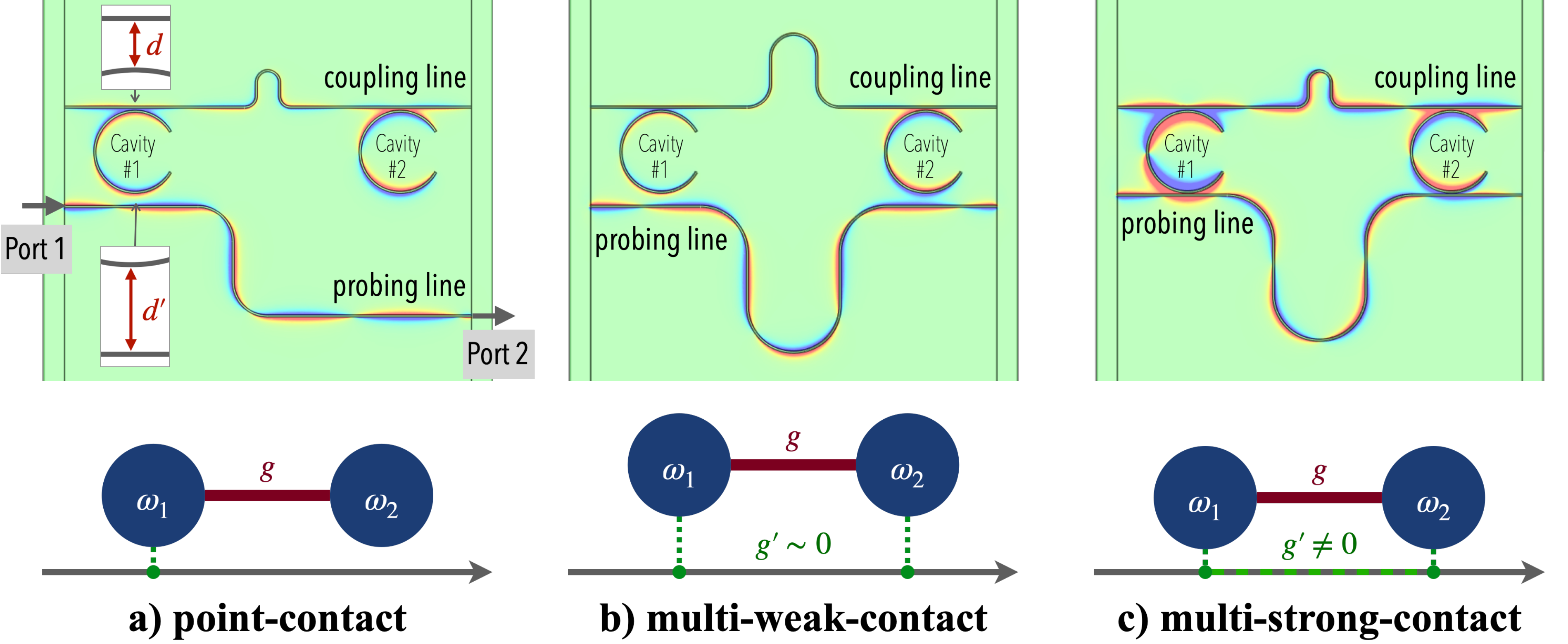}
  \caption{Two C-shaped micro-ring cavities coupled by the coupling line and detected via the probing line. The coupling strength between the cavities and the coupling/probing lines are controlled by the spacing $d$ and $d'$. a) The point-contact, the probing line couples to one cavity only. b) The multi-weak-contact, the probing line couples to  both cavities but with relatively large spacing $d' \gg d$. c) The multi-strong-contact, similar to b) but with $d' \sim d$.}
  \label{fig:fig2_system}
\end{figure}

\begin{figure*}[ht]
  \includegraphics[width=\textwidth]{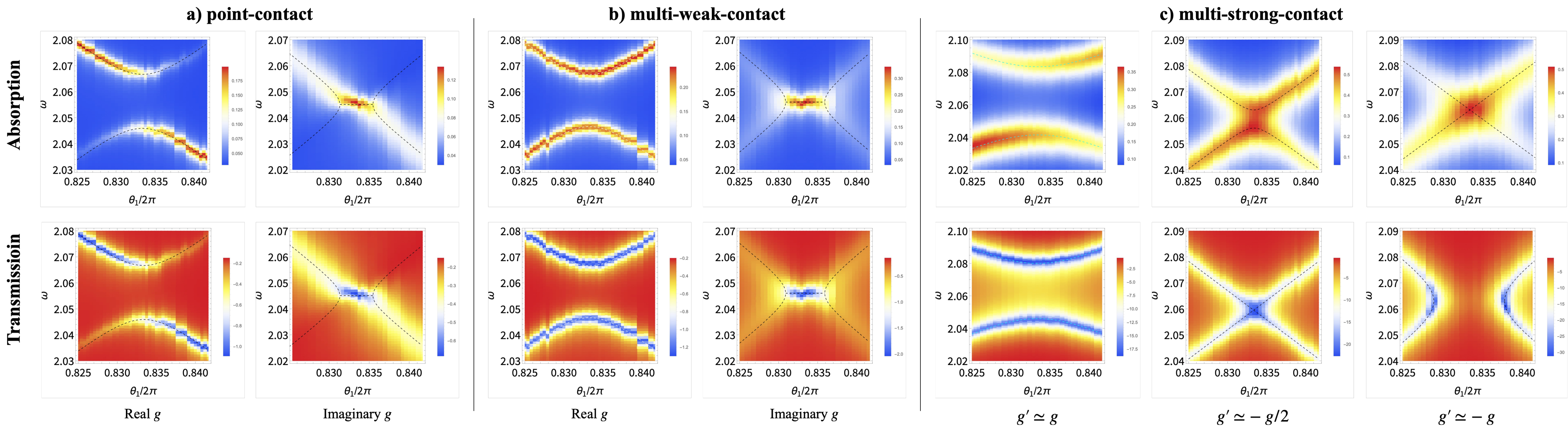} 
  \caption{a) The transmission spectra agree with the absorption spectra for both real and imaginary $g$ for the point-contact probing. b) Sam as a) for the multi-weak-contact scenario. c) The transmission spectra disagree with the absorption spectrum for the multi-strong-contact probing. }
  \label{fig:sim}
\end{figure*}

\emph{Simulation -} To illustrate the (mis)alignment between the transmission spectrum and the absorption spectrum, we consider a micro-ring system as shown in \Figure{fig:fig2_system}(a). The system consists of two C-shaped micro-ring cavities, coupled via a microstrip placed on the top. The bottom microstrip serves as the probing channel. 
The strength of direct coupling $g$ is controlled by the spacing $d$ between the coupling line and the C-cavities. The direct coupling $g$ is coherent when the ends of the coupling channel are closed. When ends of the coupling channel are open, the direct coupling $g$ can tune to be either purely real or imaginary, or even complex (see \Figure{fig:tuning} below).
The strength of the probing is controlled by the spacing $d'$ between the probing line and the C-cavities. 
%When the stripes have open ends, the phase of the coupling $g$ or $g'$ is determined by the distance between the contact points with the C-cavities.
The width of both the microstrip lines and the C-cavities is \SI{0.5}{\mu m}, and each C-cavity is portion of a full circle with a radius of \SI{10}{\mu m}. The dielectric silicon substrate beneath the microstrip lines and the cavities has thickness of \SI{0.762}{\mu m}. 
We focus on the lowest standing-wave modes of the C-cavities, whose frequency depends on total length or the opening angle of the C-cavity, denoted by $\theta_{1,2}$. When $\theta = 5\pi/3$, the lowest standing-wave mode for an isolated C-cavity has a frequency of \SI{2.04}{THz} and wavelength $\lambda \simeq \SI{52}{\mu m}$.  
The C-cavity, rather than a circular cavity, is used here to avoid the degenerate modes.
The transmission and absorption spectra of the system are simulated using the COMSOL Multiphysics \cite{COMSOL}, and \Figure{fig:fig2_system}(b) shows the simulated field distribution in the structure. 

% \section{Misalignment between the transmission \& absorption spectrum}

% \begin{figure*}[t]
%   \includegraphics[width=\textwidth]{fig2}
%   \caption{a-c): Point-contact detection $S_{21}$ spectrum for a point contact setup with real and imaginary $g$ (colormap, calculated from \Eq{eqn:$S_{21}$}) ($\gamma_1 \neq 0, \gamma_2 = 0$). d-g) multi-contact detection: The absorption and $S_{21}$ spectrum for $g'\sim 0$, $g' = g$ and $g' -g$. The green solid curves are the spectrum of the system, and the dashed curves are the spectrum of the extended system.}
%   \label{fig:fig2}
% \end{figure*}

%\subsection{Point-contact detection}

\emph{Point-contact -} We now consider the probing channel interacts exclusively with one of the C-cavities, so the coupling via the probing channel vanishes ($g' = 0$). In this case, $g_S = g_A = g$, the transmission spectrum and absorption spectrum are all identical. \Figure{fig:sim}(a) showcases the agreement between the absorption spectrum and the transmission spectrum for both real and imaginary $g$, being level repulsion and level attraction. Such alignment between the transmission and absorption spectrum is not affected by the contact strength, be it weak or strong. The frequency $\omega_{1,2}$ are tuned by varying the opening angle $\theta_{1,2}$ and keeping $(\theta_1+\theta_2)/2 = 5\pi/3$. The real and imaginary $g$ are realized by making the two ends of the coupling channel closed or open. 
Both spectra show more prominent signatures for mode-1 because the probing channel is directly in contact with mode-1, but not with mode-2. 

%\subsection{Multi-weak-contact detection}

\emph{Multi-contact -} 
We now consider the probing channel interacts with both C-cavities. Unlike the point-contact case, for the multi-contact case, the alignment between the transmission and absorption spectra relies on the strength of the probing contacts.
\Figure{fig:sim}(b) shows the case with weak contacts so that $g' \ll g$. The weak probing is realized by placing the probing channel further away from the C-cavities ($d' = \SI{3}{\mu m}$), but keeping the coupling channel close to the C-cavities ($d = \SI{0.5}{\mu m}$). In this case, \Figure{fig:sim}(b) shows that the transmission spectrum and the absorption spectrum also agree with one another perfectly. The spectrum is more balanced in the multi-weak-contact case than the point-contact case because the probing channel is directly in contact with both modes.
\Figure{fig:sim}(b) confirms that the transmission spectrum can reflect the system resonances faithfully for both point-contact and the multi-weak contact cases. 

%\subsection{Multi-strong-contact detection}

% \begin{figure}[t]
%   \includegraphics[width=\columnwidth]{fig_msc} 
%   \caption{Left: Point contact. Right: Multi-weak-contact. }
%   \label{fig:msc}
% \end{figure}

\begin{center}
\begin{table}[b]
\begin{tabular}{c|c|c|c|c} 
  $g'$ & system & extended sys. & transmission: $g_S$ & absorption: $g_A$ \\ \hline
%  $0$ & $g$ & $g$ & $g$ & $g$ \\
  $+g$ & $g$ & $2g$ & $\sqrt{3}g$ & $2g$ \\
  $-g/2$ & $g$ & $g/2$ & $0$ & $g/2$ \\
  $-g$ & $g$ & $0$ & $ig$ & $0$ 
\end{tabular}
\label{tab:gg}
\caption{When the indirect coupling is not negligible in the strong multi-contact detection case, the transmission spectrum does not faithfully reflect the system or extended system resonances.}
\end{table}
\end{center}

The interesting case happens when the probing channel is strongly coupled to both modes. The strong coupling is realized by placing the probing line close to the C-cavities with $d' = \SI{0.3}{\mu m}$, and the spacing between the coupling line (with closed ends) and the cavities is $d = \SI{0.5}{\mu m}$. 
%\wjj{Due to the coupling channels being closed on both sides, when $d = d'$, $g > g'$, so here we used $d > d'$ setup.}
As a result, the C-cavities are coupled to the coupling line and the probing line with similar strength so that $g' \sim g$. The direct coupling $g$ is real because the ends of the coupling channel are closed. The relative phase between $g$ and $g'$ can be tuned by varying the extra length by the U-shape in the middle in the probing line. For this multi-strong contact scenario, we discuss three special cases: i) $g' \simeq g$, the indirect coupling has the same strength and phase as the direct coupling; ii) $g' \simeq - g/2$, the indirect coupling has opposite phase as the direction coupling, but half the strength; iii) $g' \simeq -g$, the indirect coupling has the same strength but opposite phase as the direct coupling. \Table{tab:gg} lists the corresponding effective coupling for the isolated system without the probing line ($g$), the extended system with the probing line ($g+g'$), the transmission spectrum ($g_S$), and the absorption spectrum ($g_A$), respectively.
Apparently, $g_S$ and $g_A$ do not match with each other in all three cases. Such misalignment between the transmission and absorption spectra is confirmed in simulation shown in \Figure{fig:sim}(c): i) for $g' \simeq g$, both the absorption and the transmission spectrum show level repulsive behavior, but the absorption spectrum has larger gap than the transmission spectrum, which is qualitatively in agreement with the theoretical values $g_A = 2g > \sqrt{3}g = g_S$; ii) for $g' \simeq -g/2$, the absorption spectrum is level repulsive with a small gap, but the transmission spectrum has no coupling at all, also in agreement with the theoretical expectation of $g_A = g/2, g_S= 0$; iii) for $g' \simeq -g$, the transmission spectrum  shows a level attraction, but in fact there is no coupling in the absorption spectrum, in line with the theoretical values $g_S = i g, g_A = 0$. These results show that for the multi-strong-contact scenarios, the transmission spectrum $S_{21}$ cannot reflect the real resonance behavior of the system under detection. 

%\section{Coupling tuning: Repulsive and Attractive}

\begin{figure}[t]
  \includegraphics[width=\columnwidth]{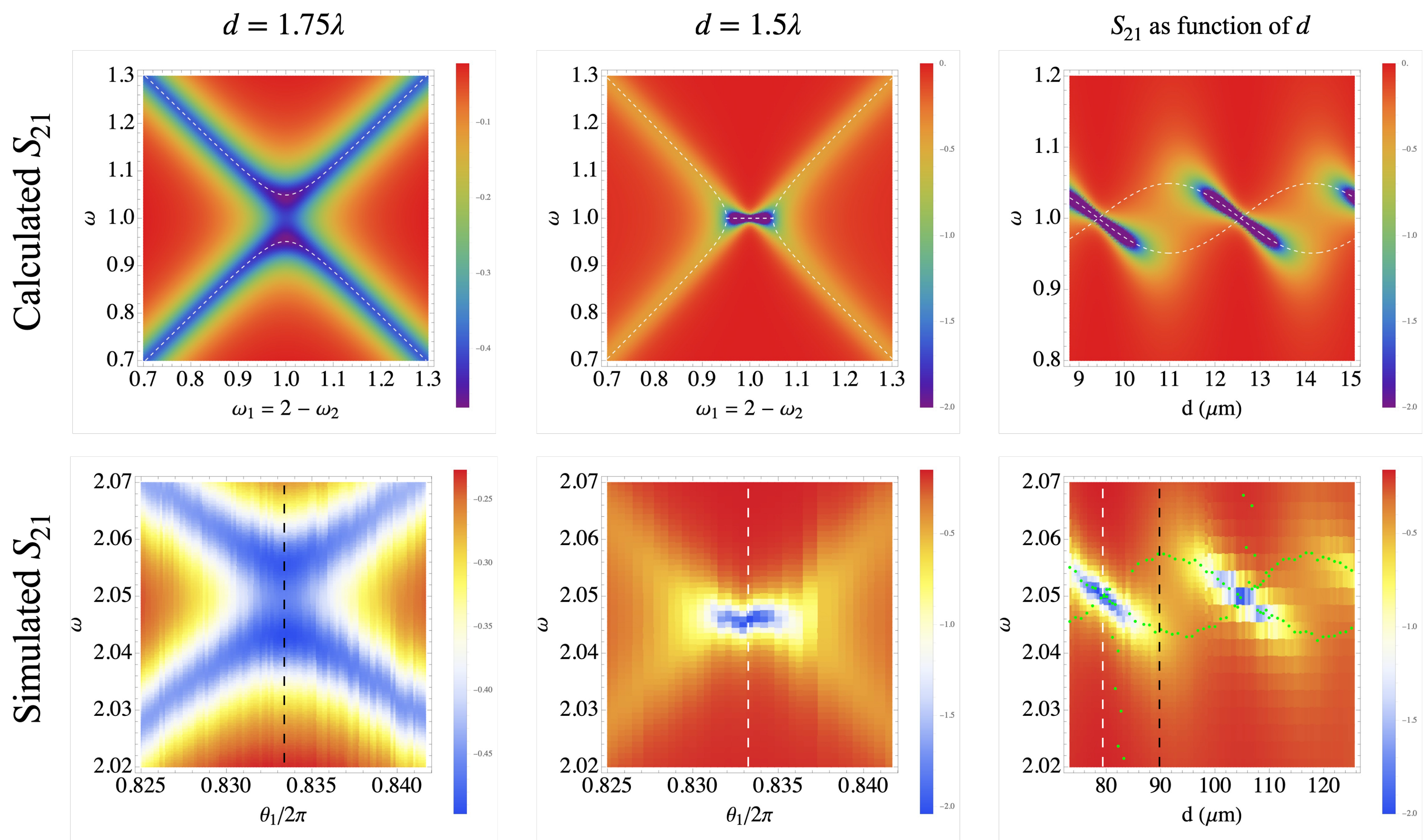}
  \caption{Tuning the spectrum by varying the length of the coupling stripe between the contacting points using multi-contact weak detection setup. We can directly observe a) level repulsion and b) level attraction from the S-matrix spectrum. The green dots in c) are the eigenfrequencies obtained from the eigenvalue solver in COMSOL simulation.}
  \label{fig:tuning}
\end{figure}

\emph{Coupling tuning by an open channel -} For the point- or multi-weak-contact cases, when the ends of the coupling channel are closed, the direct coupling $g$ is of the coherent type, thus manifesting a level repulsive spectrum. However, when the coupling channel has open ends, the direct coupling $g = -i\sqrt{\gamma_{1c}\gamma_{2c}}e^{i2\pi l/\lambda}$ is in general complex, where $\gamma_{ic}$ denotes the coupling strength between the C-cavities and the coupling stripe and $l$ is length of the coupling stripe between the contacting points. Because of the phase factor $e^{i2\pi l/\lambda}$, it is possible to tune the nature of the coupling, from real to purely imaginary. Correspondingly, the spectrum changes from repulsive to attractive. Here we demonstrate this tunability in \Figure{fig:tuning}. To avoid the complications from the probing line, we use the weak detection setup with $d' = 6 d = \SI{3}{\mu m}$, for which $g'\sim 0$ and $g_S\simeq g_A\simeq g$, therefore the transmission spectrum provides genuine information about the system under investigation. 
\Figure{fig:tuning}(a,b) show the calculated (\Eq{eqn:S21}) and simulated transmission spectrum for $l = 1.75\lambda$ and $l = 1.5\lambda$, corresponding to a coherent and dissipative coupling, respectively. This demonstrates the free propagation along an open coupling channel can be used for tuning the nature of coupling. 
\Figure{fig:tuning}(c) presents the line cut of the spectrum at zero detuning point $\omega_2 = \omega_1$ as function of length $l$, which shows an alternating repulsive and attractive behavior, with a period of $\lambda/2$ as expected.

%\section{Discussion \& Conclusion}

In conslusion, we found that the accuracy of $S_{21}$ spectra depends on the probing technique used. Point or weak probes preserve the integrity of the $S_{21}$ spectrum, while strong and multi-point probes can distort it, leading to potential errors. This underscores the importance of choosing the right probing methods for accurate characterization of resonant systems, especially when multiple modes interact. Our findings offer crucial guidance for researchers and engineers working with these systems.

\emph{Acknowledgements -} 
This work was supported by the National Key Research and Development Program of China (Grant no. 2022YFA1403300) and Shanghai Municipal Science and Technology Major Project (Grant No.2019SHZDZX01).

\bibliographystyle{apsrev}
\bibliography{ref_S21}

\begin{thebibliography}{29}
\expandafter\ifx\csname natexlab\endcsname\relax\def\natexlab#1{#1}\fi
\expandafter\ifx\csname bibnamefont\endcsname\relax
  \def\bibnamefont#1{#1}\fi
\expandafter\ifx\csname bibfnamefont\endcsname\relax
  \def\bibfnamefont#1{#1}\fi
\expandafter\ifx\csname citenamefont\endcsname\relax
  \def\citenamefont#1{#1}\fi
\expandafter\ifx\csname url\endcsname\relax
  \def\url#1{\texttt{#1}}\fi
\expandafter\ifx\csname urlprefix\endcsname\relax\def\urlprefix{URL }\fi
\providecommand{\bibinfo}[2]{#2}
\providecommand{\eprint}[2][]{\url{#2}}

\bibitem[{\citenamefont{Walther et~al.}(2006)\citenamefont{Walther, Varcoe,
  Englert, and Becker}}]{walther_cavity_2006}
\bibinfo{author}{\bibfnamefont{H.}~\bibnamefont{Walther}},
  \bibinfo{author}{\bibfnamefont{B.~T.~H.} \bibnamefont{Varcoe}},
  \bibinfo{author}{\bibfnamefont{B.-G.} \bibnamefont{Englert}},
  \bibnamefont{and} \bibinfo{author}{\bibfnamefont{T.}~\bibnamefont{Becker}},
  \bibinfo{journal}{Rep. Prog. Phys.} \textbf{\bibinfo{volume}{69}},
  \bibinfo{pages}{1325} (\bibinfo{year}{2006}).

\bibitem[{\citenamefont{Xu et~al.}(2023)\citenamefont{Xu, Gu, Li, Weng, Wang,
  Li, Wang, Zhu, and You}}]{xu_quantum_2023}
\bibinfo{author}{\bibfnamefont{D.}~\bibnamefont{Xu}},
  \bibinfo{author}{\bibfnamefont{X.-K.} \bibnamefont{Gu}},
  \bibinfo{author}{\bibfnamefont{H.-K.} \bibnamefont{Li}},
  \bibinfo{author}{\bibfnamefont{Y.-C.} \bibnamefont{Weng}},
  \bibinfo{author}{\bibfnamefont{Y.-P.} \bibnamefont{Wang}},
  \bibinfo{author}{\bibfnamefont{J.}~\bibnamefont{Li}},
  \bibinfo{author}{\bibfnamefont{H.}~\bibnamefont{Wang}},
  \bibinfo{author}{\bibfnamefont{S.-Y.} \bibnamefont{Zhu}}, \bibnamefont{and}
  \bibinfo{author}{\bibfnamefont{J.}~\bibnamefont{You}},
  \bibinfo{journal}{Physical Review Letters} \textbf{\bibinfo{volume}{130}},
  \bibinfo{pages}{193603} (\bibinfo{year}{2023}).

\bibitem[{\citenamefont{Fan}(2002)}]{fan_sharp_2002}
\bibinfo{author}{\bibfnamefont{S.}~\bibnamefont{Fan}},
  \bibinfo{journal}{Applied Physics Letters} \textbf{\bibinfo{volume}{80}},
  \bibinfo{pages}{908} (\bibinfo{year}{2002}).

\bibitem[{\citenamefont{Maleki et~al.}(2004)\citenamefont{Maleki, Matsko,
  Savchenkov, and Ilchenko}}]{maleki_tunable_2004}
\bibinfo{author}{\bibfnamefont{L.}~\bibnamefont{Maleki}},
  \bibinfo{author}{\bibfnamefont{A.~B.} \bibnamefont{Matsko}},
  \bibinfo{author}{\bibfnamefont{A.~A.} \bibnamefont{Savchenkov}},
  \bibnamefont{and} \bibinfo{author}{\bibfnamefont{V.~S.}
  \bibnamefont{Ilchenko}}, \bibinfo{journal}{Optics Letters}
  \textbf{\bibinfo{volume}{29}}, \bibinfo{pages}{626} (\bibinfo{year}{2004}).

\bibitem[{\citenamefont{Aspelmeyer et~al.}(2014)\citenamefont{Aspelmeyer,
  Kippenberg, and Marquardt}}]{aspelmeyer_cavity_2014}
\bibinfo{author}{\bibfnamefont{M.}~\bibnamefont{Aspelmeyer}},
  \bibinfo{author}{\bibfnamefont{T.~J.} \bibnamefont{Kippenberg}},
  \bibnamefont{and}
  \bibinfo{author}{\bibfnamefont{F.}~\bibnamefont{Marquardt}},
  \bibinfo{journal}{Rev. Mod. Phys.} \textbf{\bibinfo{volume}{86}},
  \bibinfo{pages}{1391} (\bibinfo{year}{2014}).

\bibitem[{\citenamefont{Lachance-Quirion
  et~al.}(2019)\citenamefont{Lachance-Quirion, Tabuchi, Gloppe, Usami, and
  Nakamura}}]{lachance-quirion_hybrid_2019}
\bibinfo{author}{\bibfnamefont{D.}~\bibnamefont{Lachance-Quirion}},
  \bibinfo{author}{\bibfnamefont{Y.}~\bibnamefont{Tabuchi}},
  \bibinfo{author}{\bibfnamefont{A.}~\bibnamefont{Gloppe}},
  \bibinfo{author}{\bibfnamefont{K.}~\bibnamefont{Usami}}, \bibnamefont{and}
  \bibinfo{author}{\bibfnamefont{Y.}~\bibnamefont{Nakamura}},
  \bibinfo{journal}{Appl. Phys. Express} \textbf{\bibinfo{volume}{12}},
  \bibinfo{pages}{070101} (\bibinfo{year}{2019}).

\bibitem[{\citenamefont{Zare~Rameshti et~al.}(2022)\citenamefont{Zare~Rameshti,
  Viola~Kusminskiy, Haigh, Usami, Lachance-Quirion, Nakamura, Hu, Tang, Bauer,
  and Blanter}}]{zare_rameshti_cavity_2022}
\bibinfo{author}{\bibfnamefont{B.}~\bibnamefont{Zare~Rameshti}},
  \bibinfo{author}{\bibfnamefont{S.}~\bibnamefont{Viola~Kusminskiy}},
  \bibinfo{author}{\bibfnamefont{J.~A.} \bibnamefont{Haigh}},
  \bibinfo{author}{\bibfnamefont{K.}~\bibnamefont{Usami}},
  \bibinfo{author}{\bibfnamefont{D.}~\bibnamefont{Lachance-Quirion}},
  \bibinfo{author}{\bibfnamefont{Y.}~\bibnamefont{Nakamura}},
  \bibinfo{author}{\bibfnamefont{C.-M.} \bibnamefont{Hu}},
  \bibinfo{author}{\bibfnamefont{H.~X.} \bibnamefont{Tang}},
  \bibinfo{author}{\bibfnamefont{G.~E.~W.} \bibnamefont{Bauer}},
  \bibnamefont{and} \bibinfo{author}{\bibfnamefont{Y.~M.}
  \bibnamefont{Blanter}}, \bibinfo{journal}{Physics Reports}
  \textbf{\bibinfo{volume}{979}}, \bibinfo{pages}{1} (\bibinfo{year}{2022}).

\bibitem[{\citenamefont{Auld}(1963)}]{auld_coupling_1963}
\bibinfo{author}{\bibfnamefont{B.~A.} \bibnamefont{Auld}},
  \bibinfo{journal}{Journal of Applied Physics} \textbf{\bibinfo{volume}{34}},
  \bibinfo{pages}{1629} (\bibinfo{year}{1963}).

\bibitem[{\citenamefont{Zhang et~al.}(2014)\citenamefont{Zhang, Zou, Jiang, and
  Tang}}]{zhang_strongly_2014}
\bibinfo{author}{\bibfnamefont{X.}~\bibnamefont{Zhang}},
  \bibinfo{author}{\bibfnamefont{C.-L.} \bibnamefont{Zou}},
  \bibinfo{author}{\bibfnamefont{L.}~\bibnamefont{Jiang}}, \bibnamefont{and}
  \bibinfo{author}{\bibfnamefont{H.~X.} \bibnamefont{Tang}},
  \bibinfo{journal}{Phys. Rev. Lett.} \textbf{\bibinfo{volume}{113}},
  \bibinfo{pages}{156401} (\bibinfo{year}{2014}).

\bibitem[{\citenamefont{Wang et~al.}(2018)\citenamefont{Wang, Zhang, Zhang, Li,
  Hu, and You}}]{wang_bistability_2018}
\bibinfo{author}{\bibfnamefont{Y.-P.} \bibnamefont{Wang}},
  \bibinfo{author}{\bibfnamefont{G.-Q.} \bibnamefont{Zhang}},
  \bibinfo{author}{\bibfnamefont{D.}~\bibnamefont{Zhang}},
  \bibinfo{author}{\bibfnamefont{T.-F.} \bibnamefont{Li}},
  \bibinfo{author}{\bibfnamefont{C.-M.} \bibnamefont{Hu}}, \bibnamefont{and}
  \bibinfo{author}{\bibfnamefont{J.}~\bibnamefont{You}},
  \bibinfo{journal}{Physical Review Letters} \textbf{\bibinfo{volume}{120}},
  \bibinfo{pages}{057202} (\bibinfo{year}{2018}).

\bibitem[{\citenamefont{Yang et~al.}(2017)\citenamefont{Yang, Liu, and
  You}}]{yang_anti-_2017}
\bibinfo{author}{\bibfnamefont{F.}~\bibnamefont{Yang}},
  \bibinfo{author}{\bibfnamefont{Y.-C.} \bibnamefont{Liu}}, \bibnamefont{and}
  \bibinfo{author}{\bibfnamefont{L.}~\bibnamefont{You}},
  \bibinfo{journal}{Physical Review A} \textbf{\bibinfo{volume}{96}},
  \bibinfo{pages}{053845} (\bibinfo{year}{2017}).

\bibitem[{\citenamefont{Wang and Hu}(2020)}]{wang_dissipative_2020}
\bibinfo{author}{\bibfnamefont{Y.-P.} \bibnamefont{Wang}} \bibnamefont{and}
  \bibinfo{author}{\bibfnamefont{C.-M.} \bibnamefont{Hu}},
  \bibinfo{journal}{Journal of Applied Physics} \textbf{\bibinfo{volume}{127}},
  \bibinfo{pages}{130901} (\bibinfo{year}{2020}).

\bibitem[{\citenamefont{Yang et~al.}(2019)\citenamefont{Yang, Rao, Gui, Yao,
  Lu, and Hu}}]{yang_control_2019}
\bibinfo{author}{\bibfnamefont{Y.}~\bibnamefont{Yang}},
  \bibinfo{author}{\bibfnamefont{J.}~\bibnamefont{Rao}},
  \bibinfo{author}{\bibfnamefont{Y.}~\bibnamefont{Gui}},
  \bibinfo{author}{\bibfnamefont{B.}~\bibnamefont{Yao}},
  \bibinfo{author}{\bibfnamefont{W.}~\bibnamefont{Lu}}, \bibnamefont{and}
  \bibinfo{author}{\bibfnamefont{C.-M.} \bibnamefont{Hu}},
  \bibinfo{journal}{Physical Review Applied} \textbf{\bibinfo{volume}{11}},
  \bibinfo{pages}{054023} (\bibinfo{year}{2019}).

\bibitem[{\citenamefont{Li et~al.}(2022)\citenamefont{Li, Yefremenko,
  Lisovenko, Trevillian, Polakovic, Cecil, Barry, Pearson, Divan, Tyberkevych
  et~al.}}]{li_coherent_2022}
\bibinfo{author}{\bibfnamefont{Y.}~\bibnamefont{Li}},
  \bibinfo{author}{\bibfnamefont{V.~G.} \bibnamefont{Yefremenko}},
  \bibinfo{author}{\bibfnamefont{M.}~\bibnamefont{Lisovenko}},
  \bibinfo{author}{\bibfnamefont{C.}~\bibnamefont{Trevillian}},
  \bibinfo{author}{\bibfnamefont{T.}~\bibnamefont{Polakovic}},
  \bibinfo{author}{\bibfnamefont{T.~W.} \bibnamefont{Cecil}},
  \bibinfo{author}{\bibfnamefont{P.~S.} \bibnamefont{Barry}},
  \bibinfo{author}{\bibfnamefont{J.}~\bibnamefont{Pearson}},
  \bibinfo{author}{\bibfnamefont{R.}~\bibnamefont{Divan}},
  \bibinfo{author}{\bibfnamefont{V.}~\bibnamefont{Tyberkevych}},
  \bibnamefont{et~al.}, \bibinfo{journal}{Physical Review Letters}
  \textbf{\bibinfo{volume}{128}}, \bibinfo{pages}{047701}
  (\bibinfo{year}{2022}).

\bibitem[{\citenamefont{Zhang et~al.}(2017)\citenamefont{Zhang, Luo, Wang, Li,
  and You}}]{zhang_observation_2017}
\bibinfo{author}{\bibfnamefont{D.}~\bibnamefont{Zhang}},
  \bibinfo{author}{\bibfnamefont{X.-Q.} \bibnamefont{Luo}},
  \bibinfo{author}{\bibfnamefont{Y.-P.} \bibnamefont{Wang}},
  \bibinfo{author}{\bibfnamefont{T.-F.} \bibnamefont{Li}}, \bibnamefont{and}
  \bibinfo{author}{\bibfnamefont{J.~Q.} \bibnamefont{You}},
  \bibinfo{journal}{Nature Communications} \textbf{\bibinfo{volume}{8}},
  \bibinfo{pages}{1368} (\bibinfo{year}{2017}), \bibinfo{note}{publisher:
  Nature Publishing Group}.

\bibitem[{\citenamefont{Lai et~al.}(2019)\citenamefont{Lai, Lu, Suh, Yuan, and
  Vahala}}]{lai_observation_2019}
\bibinfo{author}{\bibfnamefont{Y.-H.} \bibnamefont{Lai}},
  \bibinfo{author}{\bibfnamefont{Y.-K.} \bibnamefont{Lu}},
  \bibinfo{author}{\bibfnamefont{M.-G.} \bibnamefont{Suh}},
  \bibinfo{author}{\bibfnamefont{Z.}~\bibnamefont{Yuan}}, \bibnamefont{and}
  \bibinfo{author}{\bibfnamefont{K.}~\bibnamefont{Vahala}},
  \bibinfo{journal}{Nature} \textbf{\bibinfo{volume}{576}}, \bibinfo{pages}{65}
  (\bibinfo{year}{2019}).

\bibitem[{\citenamefont{Zhang et~al.}(2024)\citenamefont{Zhang, Dong, Wang, and
  Huang}}]{zhang_exceptional_2024}
\bibinfo{author}{\bibfnamefont{M.-N.} \bibnamefont{Zhang}},
  \bibinfo{author}{\bibfnamefont{L.}~\bibnamefont{Dong}},
  \bibinfo{author}{\bibfnamefont{L.-F.} \bibnamefont{Wang}}, \bibnamefont{and}
  \bibinfo{author}{\bibfnamefont{Q.-A.} \bibnamefont{Huang}},
  \bibinfo{journal}{Microsystems \& Nanoengineering}
  \textbf{\bibinfo{volume}{10}}, \bibinfo{pages}{12} (\bibinfo{year}{2024}).

\bibitem[{\citenamefont{Boller et~al.}(1991)\citenamefont{Boller, Imamo\v{g}lu,
  and Harris}}]{boller_observation_1991}
\bibinfo{author}{\bibfnamefont{K.-J.} \bibnamefont{Boller}},
  \bibinfo{author}{\bibfnamefont{A.}~\bibnamefont{Imamo\v{g}lu}},
  \bibnamefont{and} \bibinfo{author}{\bibfnamefont{S.~E.}
  \bibnamefont{Harris}}, \bibinfo{journal}{Physical Review Letters}
  \textbf{\bibinfo{volume}{66}}, \bibinfo{pages}{2593} (\bibinfo{year}{1991}).

\bibitem[{\citenamefont{Artoni}(2016)}]{artoni_electromagnetic_2016}
\bibinfo{author}{\bibfnamefont{M.}~\bibnamefont{Artoni}}, in
  \emph{\bibinfo{booktitle}{Reference {Module} in {Materials} {Science} and
  {Materials} {Engineering}}} (\bibinfo{publisher}{Elsevier},
  \bibinfo{year}{2016}), p. \bibinfo{pages}{B9780128035818011784}, ISBN
  \bibinfo{isbn}{978-0-12-803581-8}.

\bibitem[{\citenamefont{Qian et~al.}(2023)\citenamefont{Qian, Meng, Rao, Rao,
  An, Gui, and Hu}}]{qian_non-hermitian_2023}
\bibinfo{author}{\bibfnamefont{J.}~\bibnamefont{Qian}},
  \bibinfo{author}{\bibfnamefont{C.~H.} \bibnamefont{Meng}},
  \bibinfo{author}{\bibfnamefont{J.~W.} \bibnamefont{Rao}},
  \bibinfo{author}{\bibfnamefont{Z.~J.} \bibnamefont{Rao}},
  \bibinfo{author}{\bibfnamefont{Z.}~\bibnamefont{An}},
  \bibinfo{author}{\bibfnamefont{Y.}~\bibnamefont{Gui}}, \bibnamefont{and}
  \bibinfo{author}{\bibfnamefont{C.~M.} \bibnamefont{Hu}},
  \bibinfo{journal}{Nature Communications} \textbf{\bibinfo{volume}{14}},
  \bibinfo{pages}{3437} (\bibinfo{year}{2023}).

\bibitem[{\citenamefont{Miroshnichenko
  et~al.}(2005)\citenamefont{Miroshnichenko, Mingaleev, Flach, and
  Kivshar}}]{miroshnichenko_nonlinear_2005}
\bibinfo{author}{\bibfnamefont{A.~E.} \bibnamefont{Miroshnichenko}},
  \bibinfo{author}{\bibfnamefont{S.~F.} \bibnamefont{Mingaleev}},
  \bibinfo{author}{\bibfnamefont{S.}~\bibnamefont{Flach}}, \bibnamefont{and}
  \bibinfo{author}{\bibfnamefont{Y.~S.} \bibnamefont{Kivshar}},
  \bibinfo{journal}{Physical Review E} \textbf{\bibinfo{volume}{71}},
  \bibinfo{pages}{036626} (\bibinfo{year}{2005}).

\bibitem[{\citenamefont{Miroshnichenko
  et~al.}(2010)\citenamefont{Miroshnichenko, Flach, and
  Kivshar}}]{miroshnichenko_fano_2010}
\bibinfo{author}{\bibfnamefont{A.~E.} \bibnamefont{Miroshnichenko}},
  \bibinfo{author}{\bibfnamefont{S.}~\bibnamefont{Flach}}, \bibnamefont{and}
  \bibinfo{author}{\bibfnamefont{Y.~S.} \bibnamefont{Kivshar}},
  \bibinfo{journal}{Reviews of Modern Physics} \textbf{\bibinfo{volume}{82}},
  \bibinfo{pages}{2257} (\bibinfo{year}{2010}).

\bibitem[{\citenamefont{Han et~al.}(2023)\citenamefont{Han, Meng, Pan, Qian,
  Rao, Zhu, Gui, Hu, and An}}]{han_bound_2023}
\bibinfo{author}{\bibfnamefont{Y.}~\bibnamefont{Han}},
  \bibinfo{author}{\bibfnamefont{C.}~\bibnamefont{Meng}},
  \bibinfo{author}{\bibfnamefont{H.}~\bibnamefont{Pan}},
  \bibinfo{author}{\bibfnamefont{J.}~\bibnamefont{Qian}},
  \bibinfo{author}{\bibfnamefont{Z.}~\bibnamefont{Rao}},
  \bibinfo{author}{\bibfnamefont{L.}~\bibnamefont{Zhu}},
  \bibinfo{author}{\bibfnamefont{Y.}~\bibnamefont{Gui}},
  \bibinfo{author}{\bibfnamefont{C.-M.} \bibnamefont{Hu}}, \bibnamefont{and}
  \bibinfo{author}{\bibfnamefont{Z.}~\bibnamefont{An}},
  \bibinfo{journal}{Science Advances} \textbf{\bibinfo{volume}{9}},
  \bibinfo{pages}{eadg4730} (\bibinfo{year}{2023}).

\bibitem[{\citenamefont{Yao et~al.}(2019)\citenamefont{Yao, Yu, Zhang, Lu, Gui,
  Hu, and Blanter}}]{yao_microscopic_2019}
\bibinfo{author}{\bibfnamefont{B.}~\bibnamefont{Yao}},
  \bibinfo{author}{\bibfnamefont{T.}~\bibnamefont{Yu}},
  \bibinfo{author}{\bibfnamefont{X.}~\bibnamefont{Zhang}},
  \bibinfo{author}{\bibfnamefont{W.}~\bibnamefont{Lu}},
  \bibinfo{author}{\bibfnamefont{Y.}~\bibnamefont{Gui}},
  \bibinfo{author}{\bibfnamefont{C.-M.} \bibnamefont{Hu}}, \bibnamefont{and}
  \bibinfo{author}{\bibfnamefont{Y.~M.} \bibnamefont{Blanter}},
  \bibinfo{journal}{Phys. Rev. B} \textbf{\bibinfo{volume}{100}},
  \bibinfo{pages}{214426} (\bibinfo{year}{2019}).

\bibitem[{\citenamefont{Collett and Gardiner}(1984)}]{collett_squeezing_1984}
\bibinfo{author}{\bibfnamefont{M.~J.} \bibnamefont{Collett}} \bibnamefont{and}
  \bibinfo{author}{\bibfnamefont{C.~W.} \bibnamefont{Gardiner}},
  \bibinfo{journal}{Phys. Rev. A} \textbf{\bibinfo{volume}{30}},
  \bibinfo{pages}{1386} (\bibinfo{year}{1984}).

\bibitem[{\citenamefont{Garrison and Chiao}(2008)}]{garrison_quantum_2008}
\bibinfo{author}{\bibfnamefont{J.}~\bibnamefont{Garrison}} \bibnamefont{and}
  \bibinfo{author}{\bibfnamefont{R.}~\bibnamefont{Chiao}},
  \emph{\bibinfo{title}{Quantum {Optics}}} (\bibinfo{publisher}{Oxford
  University Press}, \bibinfo{year}{2008}), ISBN
  \bibinfo{isbn}{978-0-19-170864-0}.

\bibitem[{\citenamefont{Walls and Milburn}(2008)}]{walls_quantum_2008}
\bibinfo{author}{\bibfnamefont{D.~F.} \bibnamefont{Walls}} \bibnamefont{and}
  \bibinfo{author}{\bibfnamefont{G.~J.} \bibnamefont{Milburn}},
  \emph{\bibinfo{title}{Quantum optics}} (\bibinfo{publisher}{Springer},
  \bibinfo{address}{Berlin}, \bibinfo{year}{2008}), \bibinfo{edition}{2nd} ed.,
  ISBN \bibinfo{isbn}{978-3-540-28573-1}.

\bibitem[{\citenamefont{Yuan et~al.}(2020)\citenamefont{Yuan, Yu, and
  Xiao}}]{yuan_loop_2020}
\bibinfo{author}{\bibfnamefont{H.~Y.} \bibnamefont{Yuan}},
  \bibinfo{author}{\bibfnamefont{W.}~\bibnamefont{Yu}}, \bibnamefont{and}
  \bibinfo{author}{\bibfnamefont{J.}~\bibnamefont{Xiao}},
  \bibinfo{journal}{Phys. Rev. A} \textbf{\bibinfo{volume}{101}},
  \bibinfo{pages}{043824} (\bibinfo{year}{2020}).

\bibitem[{COM(2024)}]{COMSOL}
\bibinfo{journal}{COMSOL Multiphysics \textregistered ~ v. 5.4. www.comsol.com.
  COMSOL AB, Stockholm, Sweden}  (\bibinfo{year}{2024}).

\end{thebibliography}

\appendix 

\section{Appendix: The calculation of $S_{21}$ using the loop theory}

\begin{figure}[t]
  \includegraphics[width=\columnwidth]{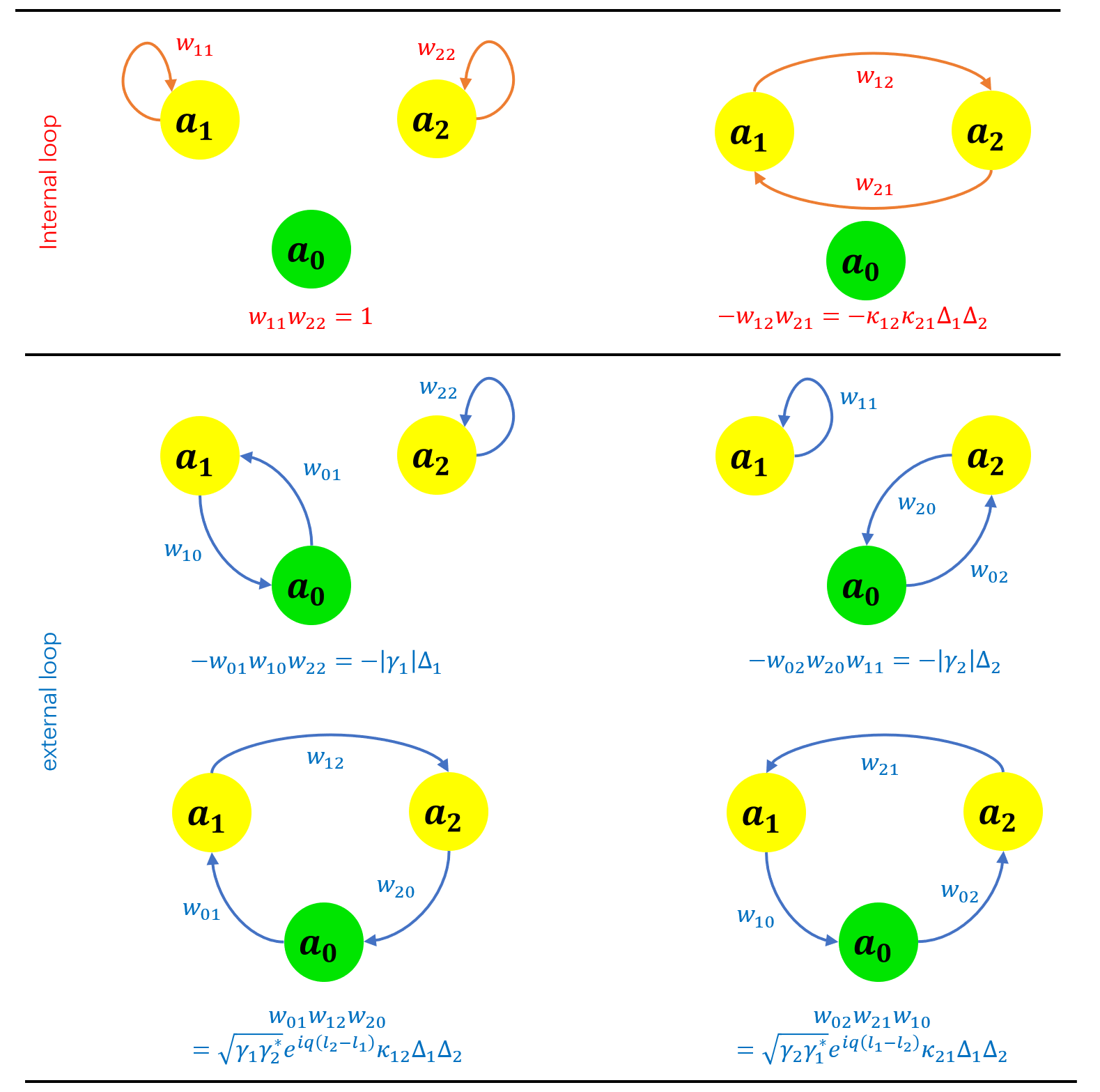}
  \caption{}
  \label{fig:loop}
\end{figure}

Loop theory developed by Yuan \etal \cite{yuan_loop_2020} states that the transmission $S_{21}$ is given by the following equation:
\begin{equation}
  \label{eqn:S21Gn}
  S_{21}(G_n) = 1 + i~\frac{-\displaystyle \sum_\ssf{external} \cA(F_{n+1})}{\displaystyle\sum_\ssf{internal} \cA(F_n)}.
\end{equation}
where $G_n$ is the graph representation of the system under investigation. For the case considered in this paper, the graph $G_n$ with $n = 2$ consists of two internal nodes (corresponding to the two modes) and one external node (corresponding to the mode in the probing channel). 
The external (internal) $F_n$ loops are graphs that each (internal) node has one incoming and one outgoing edge, as shown in \Figure{fig:loop}. Each loop is associated with an amplitude $\cA(F_m)$ defined as the product of the weights ($W(e_{ij})$) of all edges in the loop upto an overall sign $P(F_m)$:
\begin{equation}
  \label{eqn:AF}
  \cA(F_m) = P(F_m)\prod_{e_{ij}\in E(F_m)} W(e_{ij}).
\end{equation}
Here $P(F_m) = (-1)^{m-k}$ is the parity of permutation $F_m$, where $k$ is the number of disjoint cycles of $F_m$.
The weight between node-$i$ and -$j$
\begin{equation}
  \label{eqn:wij}
    W(e_{ij}) = w_{ij} = \kappa_{ij}\sqrt{\Delta_i\Delta_j}.
\end{equation}
is the product of the coupling strength $\kappa_{ij}$ between the two nodes with the detuning of the two nodes $\Delta_{i,j}$. Specifically, the self-coupling $\kappa_{ii} \equiv \Delta_i^{-1} \equiv \omega_i - \omega - i|\gamma_i|$ and the detuning of the probing mode $\Delta_0 \equiv 1$.

For the system considered in this paper, the two modes placed at $x_{1,2}$ are spatially separated, therefore the probing wave is coupled to the two modes with different phase, which means the coupling strength parameter
\begin{equation}
  \label{eqn:kappa0i}
    \kappa_{0i} = \kappa_{i0}^* = \sqrt{\gamma_i^*} ~ e^{iqx_i}.
\end{equation}
Furthermore, the two modes has both direct coupling $g_{ij}$ and indirect coupling via the probing channel
\begin{subequations}
  \label{eqn:kappaij}
\begin{align}
    \kappa_{12} &= g_{12} + g'_{12} = g_{12} - i\sqrt{\gamma_1 \gamma_2^*}~e^{iq_0(x_2-x_1)}, \\
    \kappa_{21} &= g_{21} + g'_{21} = g_{21} - i\sqrt{\gamma_2 \gamma_1^*}~e^{-iq_0(x_1-x_2)}.
\end{align}
\end{subequations}

By plugging \Eqs{eqn:kappa0i}{eqn:kappaij} into \Eqss{eqn:S21Gn}{eqn:wij}, we may directly find the transmission
%\jx{Check these two expressions of $S_{21}$ agree with one another.}
\begin{equation}
  S_{21} = 1 + i\frac{|\gamma_1|\Delta_1 + \abs{\gamma_2}\Delta_2 
  - i\qty(\kappa_{12} g'_{21} - \kappa_{21}{g'}_{21}^*)\Delta_1\Delta_2}{1 - \kappa_{12}\kappa_{21}\Delta_1\Delta_2}.
\end{equation}

% \onecolumngrid
% \add{
%   Substitute the expression ${g'}_{21} = -i\sqrt{\gamma_2^* \gamma_1} e^{iql}$ into the previous equation to obtain:
%   \begin{equation}
%     \begin{aligned}
%       S_{21} &= 1 + i\frac{|\gamma_1|\Delta_1 + \abs{\gamma_2}\Delta_2 
%   - i\qty(\kappa_{12} (-i\sqrt{\gamma_2^* \gamma_1} e^{iql}) - \kappa_{21}(i\sqrt{\gamma_2 \gamma_1^*} e^{-iql}))\Delta_1\Delta_2}{1 - \kappa_{12}\kappa_{21}\Delta_1\Delta_2}, \\
%   S_{21} &= 1 + i\frac{|\gamma_1|\Delta_1 + \abs{\gamma_2}\Delta_2 
%   - \qty(\kappa_{12}\sqrt{\gamma_2^* \gamma_1} e^{iql} + \kappa_{21}\sqrt{\gamma_2 \gamma_1^*}e^{-iql})\Delta_1\Delta_2}{1 - \kappa_{12}\kappa_{21}\Delta_1\Delta_2}.
%     \end{aligned}
%   \end{equation}
% Then we get:
% }
% \begin{equation}
%   S_{21} = 1 + i\frac{|\gamma_1|\Delta_1 + |\gamma_2|\Delta_2 - (\kappa_{12}\sqrt{\gamma_1\gamma_2^*}e^{iq(l_2 - l_1)} + \kappa_{21}\sqrt{\gamma_2\gamma_1^*}e^{-iq(l_2 - l_1)})\Delta_1\Delta_2}{1 - \kappa_{12}\kappa_{21}\Delta_1\Delta_2}
% \end{equation}

\end{document}